\documentclass[12pt, twocolumn,showpacs,preprintnumbers,aps,prl,reprint,groupedaddress]{revtex4-1}

\usepackage{graphicx}
\usepackage{subfigure}
\usepackage{color}
\usepackage{amsmath}
\usepackage[linktocpage,colorlinks=true,linkcolor=blue,citecolor=blue]{hyperref}

\begin{document}

\preprint{}

\title{Velocity Correlations In An Active Nematic}

\author{Sumesh P. Thampi}
\affiliation{The Rudolf Peierls Centre for Theoretical Physics, 1 Keble Road, Oxford, OX1 3NP, UK}

\author{Ramin Golestanian}
\affiliation{The Rudolf Peierls Centre for Theoretical Physics, 1 Keble Road, Oxford, OX1 3NP, UK}

\author{Julia M. Yeomans}
\affiliation{The Rudolf Peierls Centre for Theoretical Physics, 1 Keble Road, Oxford, OX1 3NP, UK}
\email[]{j.yeomans1@physics.ox.ac.uk}
\homepage[]{http://www-thphys.physics.ox.ac.uk/people/JuliaYeomans/}


\date{\today}

\begin{abstract}
{The flow properties of a continuum model for an active nematic is studied and compared with recent experiments on suspensions of microtubule bundles and molecular motors. The velocity correlation length is found to be independent of the strength of the activity while the characteristic velocity scale increases monotonically as the activity is increased, both in agreement with the experimental observations. We interpret our results in terms of the creation and annihilation dynamics of a gas of topological defects.}

\end{abstract}

\pacs{}


\maketitle

Active systems that produce their own energy, such as bacterial suspensions and living cells, are proving  rich research areas \cite{Ganesh2011, Marchettiarxiv}. Apart from their obvious relevance to a quantitative understanding of biological functions, active systems naturally operate out of thermodynamic equilibrium and hence provide a testing ground for theories of non-equilibrium statistical physics \cite{Sriram2010}. 
Hydrodynamic instabilities are inherent to active fluids \cite{Dogic2012, Sriram2002, Joanny2005, Madan2007, Scott2009}, and active suspensions spontaneously generate complex flow patterns at length scales much larger than the actual constituents \cite{Libchaber2000, Kessler2004, Julia2012, Goldstein2012, Mahadevan2011, Shelly2007, Shelly2012, Graham2005, Wolgemuth2008}. These are characterised by strong variations in vorticity (see Fig.~\ref{fig:defects}). Such turbulent-like patterns have been observed in experiments on mixtures of actin or microtubules and motor proteins designed to isolate the important ingredients leading to cellular motility \cite{Dogic2012, Bausch2008, Bausch2010, Chate2012}.
Very similar structured flows have been observed in two-dimensional layers of epithelial cells  \cite{Lenepreprint} and, at larger length scales, in dense suspensions of swimming bacteria \cite{Libchaber2000, Kessler2004, Julia2012}. The features of the flow have been reproduced qualitatively in simulations of driven rods \cite{Narayan2007, aparna2008} and microswimmers \cite{Graham2005, Shelly2007, Shelly2012} and in continuum theories of active nematics \cite{Mahadevan2011, Goldstein2012, Julia2012}.

\begin{figure}[htp!]
\centering
\subfigure[]{\includegraphics[trim=18 79 25 30,clip,width=\linewidth]{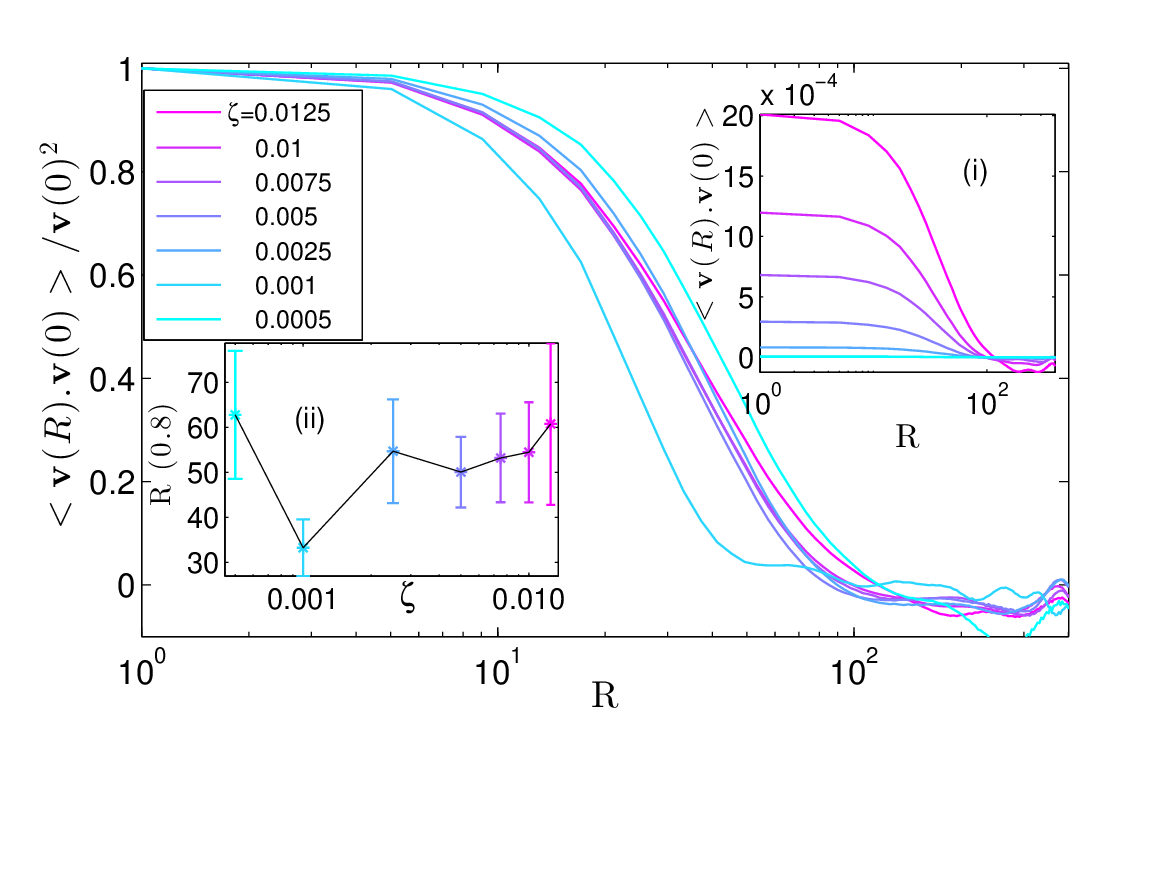}\label{fig:vcorrsim}}
\subfigure[]{\includegraphics[trim=0 0 0 0,clip,width=0.75\linewidth]{natureexpcomparelognorm.eps}\label{fig:vcorrcom}}
\caption{
(a) The velocity-velocity correlation function and (b) the RMS velocity of a 2D, active nematic suspension. In (a), normalised velocity-velocity correlations, determined by spatial and temporal averaging, are plotted as a function of distance, $R$ for different values of the activity. Unnormalised functions are shown in the inset (i). The collapse of the normalised correlation functions onto a single curve indicates insignificant correlation between the characteristic flow length scales and activity. This is also illustrated in inset (ii) where a characteristic length, defined as the value of $R$ where the velocity-velocity correlation function has fallen to 0.8 of its maximum value,
is plotted as a function of activity, $\zeta$. Fig.~\ref{fig:vcorrcom} shows a comparison between experimental \cite{Dogic2012} and simulation results for the variation of the RMS velocity with activity. Mapping of simulation to physical units for velocity is explained in the text.}
\label{fig:vcorrs}
\end{figure}

The similarity in the hydrodynamic behaviour of the different active experimental and model systems is very appealing, but  there are still large gaps in our understanding of the physical mechanisms driving active turbulence. Moreover the extent of and reasons behind any `universal' behaviour remains unclear \cite{Julia2012, Jornarxiv}. To help address these questions in this paper we measure the root mean square (RMS) velocity and velocity-velocity correlation function in a continuum model of an active suspension, as a function of the activity, and compare it to recent experiments on  extensile microtubule bundles driven by the motor protein kinesin \cite{Dogic2012}. This comparison is summarised in Fig.~\ref{fig:vcorrs}.

Figure \ref{fig:vcorrsim} shows that the normalised velocity-velocity correlation functions collapse onto a single curve, independent of the strength of the activity over a wide range of activities. The same scaling behaviour is seen in the experiments (see Fig~2(c) of Ref. \cite{Dogic2012}). This indicates that there is a length scale in the problem which is set by the structure of the underlying fluid, independent of the strength of the activity.  Figure~\ref{fig:vcorrcom}  compares the RMS velocity in the simulations and experiments where, to make the comparison, we assume a logarithmic dependence of the activity on ATP concentration. Two distinct regions are observed, a gentle increase in the velocity with activity at low activities, followed by a sharper increase at higher activities. (The simulations then become unstable and we do not see the saturation observed in the experiments.)

We will relate this behaviour to features of the velocity field, which we interpret in terms of velocity jets driven by defects and distortions of the director field \cite{Scott2009}. The length scale of velocity correlations is set by the dynamics of topological defects and depends strongly on the value of the elastic constant but only very weakly on the strength of the activity. The magnitude of the velocity, however, depends on the flow induced by the defects, which increases linearly with activity. At very low activities, individual defects are replaced by lines of kinks in the director field, which lead to smaller velocity magnitudes. We show that similar features are seen for both extensile  and contractile active nematics,  despite the coupling between the defects and the flow field being different in the two cases.

First we describe the equations of motion, those corresponding to an active nematic, that we use to model the active suspension. These are the standard equations of liquid crystal hydrodynamics, written in terms of a tensor order parameter $Q$, together with an active term which means that any gradient in $Q$ will produce a flow field.  Evolution of $\mathbf{Q}$ along with the momentum $\rho \mathbf{u}$ is given by \cite{Berisbook, DeGennesBook},
\begin{align}
(\partial_t + u_k \partial_k) Q_{ij} - S_{ij} &= \Gamma H_{ij},
\label{eqn:lc}\\
\rho (\partial_t + u_k \partial_k) u_i &= \partial_j \Pi_{ij}.
\label{eqn:ns}
\end{align}
The strain rate tensor, $E_{ij} = (\partial_i u_j + \partial_j u_i)/2$ and the vorticity tensor, $\Omega_{ij} = (\partial_j u_i - \partial_i u_j)/2$ describe the generalised advection term
$S_{ij} = (\lambda E_{ik} + \Omega_{ik})(Q_{kj} + \delta_{kj}/3) + (Q_{ik} + \delta_{ik}/3)
 (\lambda E_{kj} - \Omega_{kj}) - 2 \lambda (Q_{ij} + \delta_{ij}/3)(Q_{kl}\partial_k u_l)$,
where $\lambda$ is the alignment parameter. We choose $\lambda=0.7$ corresponding to tumbling rods \cite{Scott2009}.
  Rotational diffusivity is denoted by $\Gamma$ and the molecular field $H_{ij} = -\delta \mathcal{F}/ \delta Q_{ij} + (\delta_{ij}/3) {\rm Tr} (\delta \mathcal{F}/ \delta Q_{kl})$ is determined from the free energy, 
$\mathcal{F} = K (\partial_k Q_{ij})^2/2 + A Q_{ij} Q_{ji}/2 + B Q_{ij} Q_{jk} Q_{ki}/3 + C (Q_{ij} Q_{ji})^2/4$.
Here $K$ is the elastic constant, $A, B$ and $C$ are material constants.  The total stress generating the hydrodynamics has 3 parts; (i) the viscous stress, $\Pi_{ij}^{viscous} = 2 \mu E_{ij}$,  (ii) the passive stress,
$\Pi_{ij}^{passive}=-P\delta_{ij} + 2 \lambda(Q_{ij} + \delta_{ij}/3) (Q_{kl} H_{lk})
-\lambda H_{ik} (Q_{kj} + \delta_{kj}/3)  - \lambda (Q_{ik} + \delta_{ik}/3) H_{kj}
-\partial_i Q_{kl} \frac{\delta \mathcal{F}}{\delta \partial_j Q_{lk}} + Q_{ik}H_{kj} - H_{ik} Q_{kj}$
and the (iii) active stress, $\Pi_{ij}^{active} = -\zeta Q_{ij}$ introduced in \cite{Sriram2002}. Here $P$ is the pressure and $\zeta$ is the activity coefficient. Model details can be found in  \cite{Berisbook, DeGennesBook, Denniston2001, Denniston2004, Davide2007, Cates2008, Orlandini2008, Suzanne2011, Miha2013, Henrich2010}.

We solve this coupled system of governing equations (\ref{eqn:lc}) and (\ref{eqn:ns}) using a hybrid lattice Boltzmann method on a $D3Q15$ lattice \cite{Davide2007, Suzanne2011, Miha2013}. Simulations are performed on a $400 \times 400$ domain and the parameters used are $\Gamma=0.34$, $A=0.0$, $B=-0.3$, $C=0.3$, $K=0.02$, $\mu=2$, in lattice units unless mentioned otherwise. (Choosing $\Delta x = 2\mu$m, $\Delta t = 0.02 $s (rationalised below), these map to $\Gamma=0.0136$ (Pa s)$^{-1}$, $B=375 $Nm$^{-2}$, $\mu=0.5 $Pa\;s, if $K=1$pN \cite{Dogic2004} is used as a force scale \cite{Henrich2010}.)


It is well known that active liquid crystal equations predict hydrodynamic instabilities \cite{Sriram2002, Joanny2005, Madan2007} leading to irregular, turbulent-like flow patterns. The left column of Fig.~\ref{fig:defects} depicts a typical flow pattern observed  for an extensile suspension. The top figure is on the scale of nematic correlation lengths, the middle frame corresponds to  intermediate lengths at which individual vortices are visible, and bottom frame to hydrodynamic length scales. The region within the white border  in each figure corresponds to the figure above.  In each figure, colour shading, continuous lines and dashes correspond to the vorticity, stream lines and director field respectively.

Consider first  Fig.~\ref{fig:extdefectzoom2}. This shows a topological defect of charge $+1/2$ in the director field. The director field consequently has a strong bend deformation which leads to an active stress that generates a flow \cite{Scott2009} along the axis of symmetry of the defect, indicated by the streamlines in the figure. The vorticity field helps to locate the source of the jet, as a transition from blue, representing clockwise vorticity, to red, anticlockwise vorticity.

Moving to larger length scales, a large number of defects, and hence jets, may  be identified. The position and orientation of the defects are random, leading to jets that interact to produce complex velocity fields. For example,  Fig.~\ref{fig:extdefectzoom1} shows four $+1/2$ defects that lead to a circulating velocity field. At the larger length scale of Fig.~\ref{fig:extdefectzoom0} several such circulating flows may be seen each of which can be identified with two or more bend defects that produce  jets moving in different directions. Defects oriented in the same direction generate parallel flows and hence longer and stronger jets. Defects of charge $-1/2$ can also be identified in the field. These may be identified as a combination of three bend or splay deformations oriented such that the jets produced by each tend to cancel out. Therefore such defects act as effective stagnation points, mostly altering the flow directions of jets and hence contributing to the circulating flows.

\begin{figure}
\begin{minipage}{0.48\linewidth}
\subfigure[ ]{\includegraphics[trim=80 30 70 25, clip, width=0.9\linewidth]{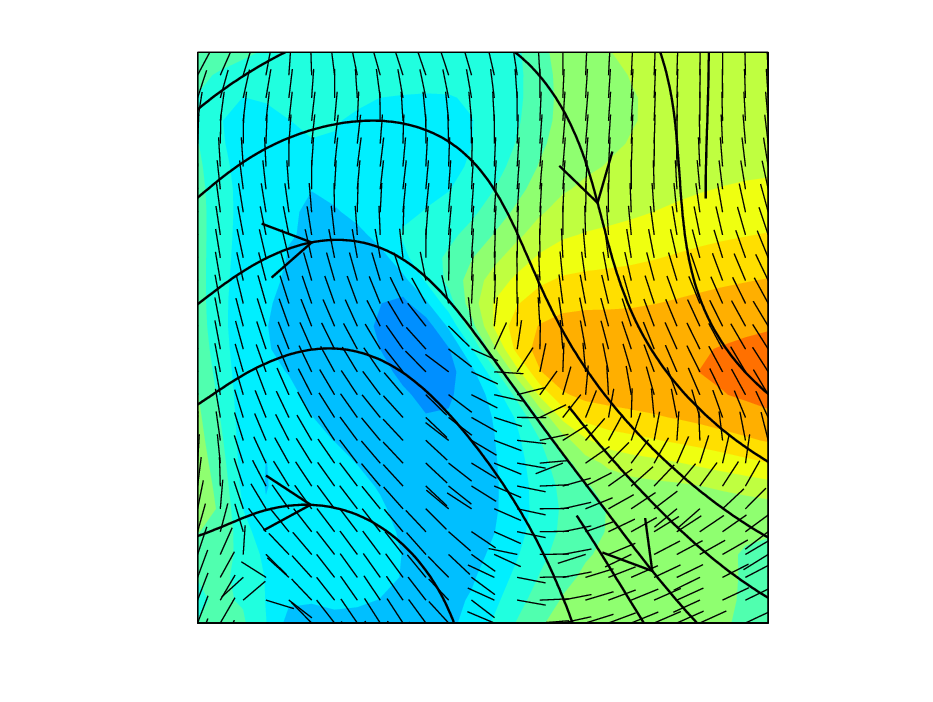}\label{fig:extdefectzoom2}}\\
\subfigure[ ]{\includegraphics[trim=80 30 70 25, clip, width=0.9\linewidth]{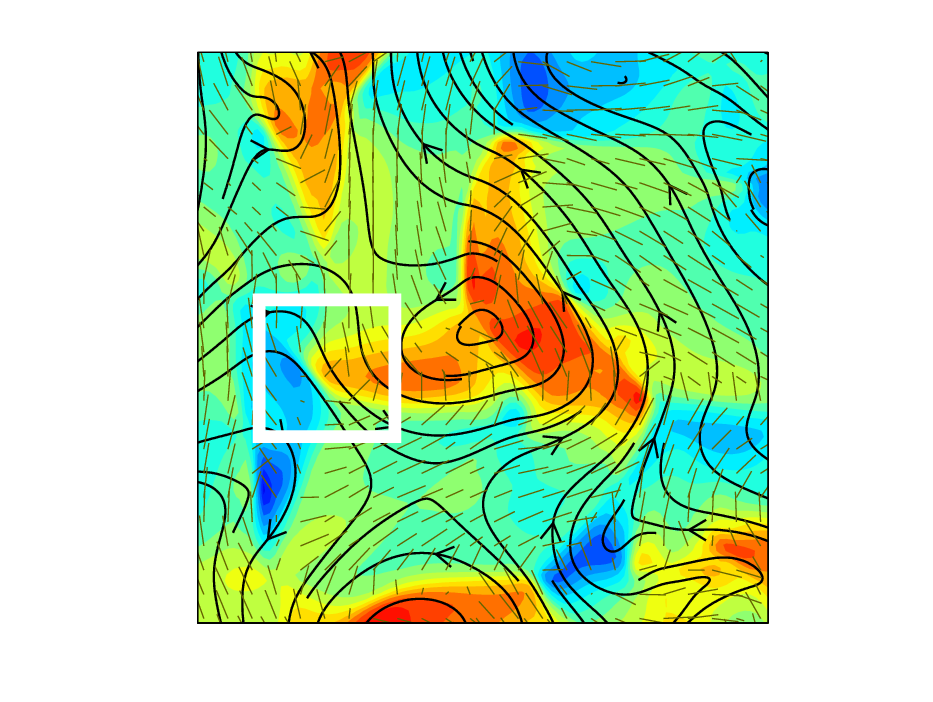}\label{fig:extdefectzoom1}}\\
\subfigure[] {\includegraphics[trim=100 30 90 20, clip, width=\linewidth]{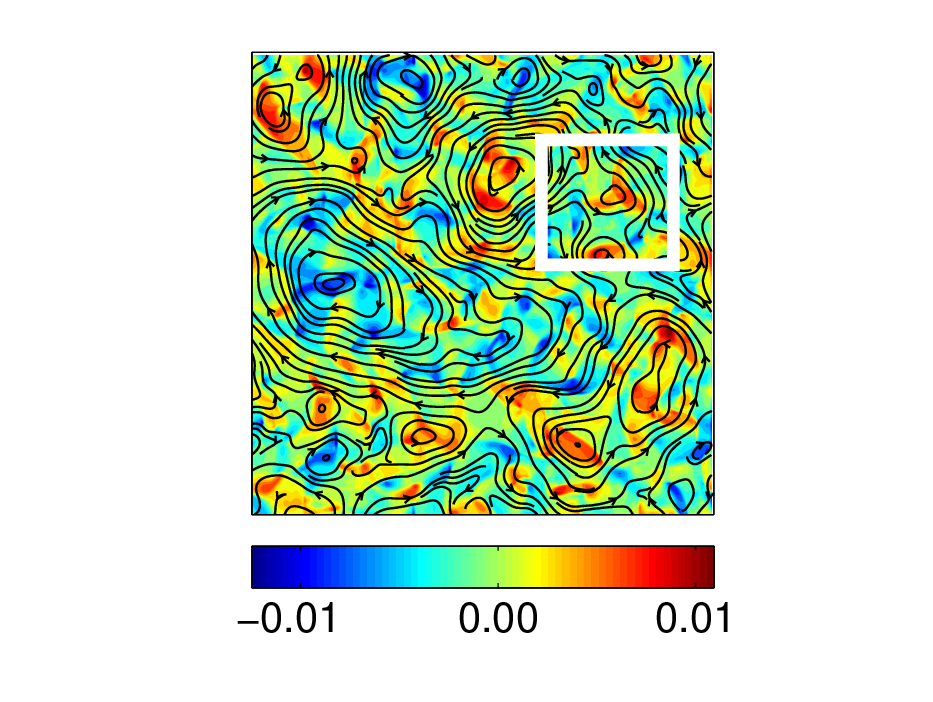}\label{fig:extdefectzoom0}}
\end{minipage}
\begin{minipage}{0.48\linewidth}
\subfigure[ ]{\includegraphics[trim=80 30 70 25, clip, width=0.9\linewidth]{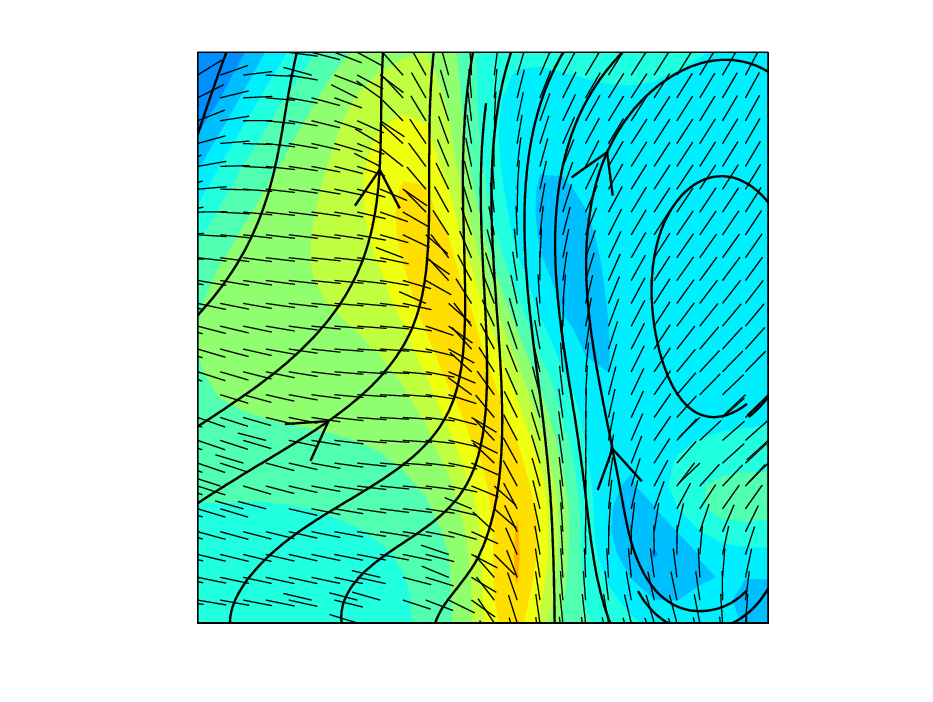}\label{fig:condefectzoom2}}\\
\subfigure[ ]{\includegraphics[trim=80 30 70 25, clip, width=0.9\linewidth]{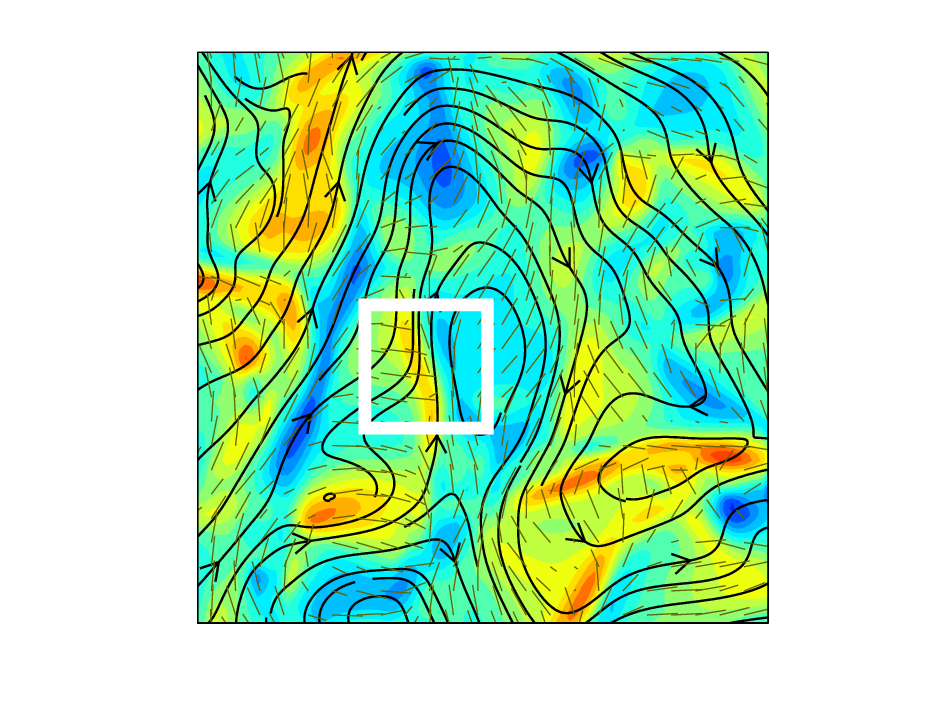}\label{fig:condefectzoom1}}\\
\subfigure[ ]{\includegraphics[trim=100 30 90 20, clip, width=\linewidth]{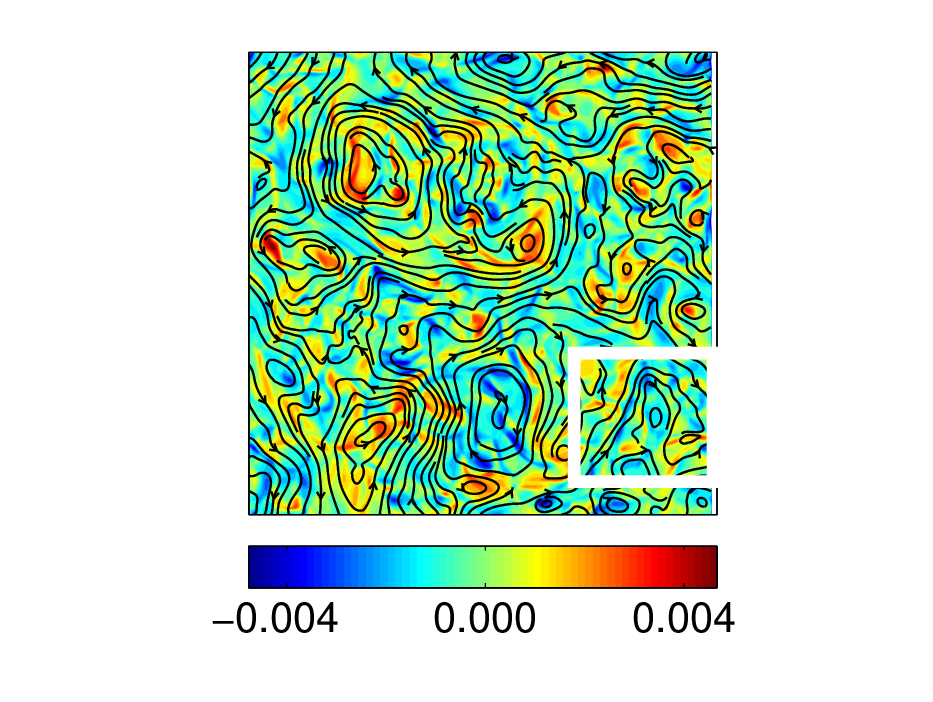}\label{fig:condefectzoom0}}
\end{minipage}
\caption{Flow and director fields at three different length scales in active nematic suspensions. The length scale increases from top to bottom, and each frame occupies the white square in the frame below. Figures in the left column (right column) correspond to extensile (contractile) stress in the suspension. Colour shading, continuous lines, and dashes  represent vorticity, stream lines and the director field respectively. Defects and deformations in the director field generate fluid jets as illustrated in Fig.~(a), (d). (Note that a defect source can be distinguished by a sharp colour change in the vorticity. Blue and red colours correspond to clockwise and anticlockwise vorticity respectively.)}
\label{fig:defects}
\end{figure}

The right hand column of Fig.~\ref{fig:defects} presents similar data for a contractile suspension. Again the flow is driven by variations in the director field. However, in the extensile case these are predominantly bend, distortions, whereas in the contractile case they are predominantly splay. This corresponds to the primary instabilities in the active nematic director field which are bend for a contractile, tumbling material and splay for an extensile one \cite{Madan2007, Scott2009}. The velocity fields are $\sim 50-70\%$ lower in the contractile suspensions for the same value of the activity. Indeed, the critical activity needed to trigger spontaneous active flow in one dimension is known to be lower for an extensile than for a contractile system \cite{Scott2009}.

\begin{figure}[h!]
\includegraphics[trim = 0 0 0 0, clip, width=\linewidth]{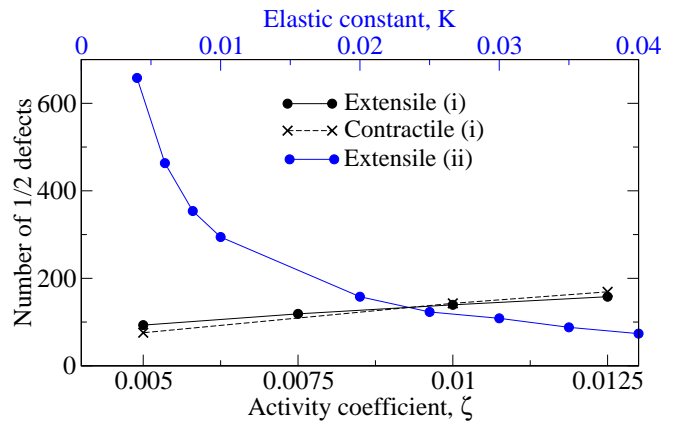}
\caption{Total number of point $1/2$ defects plotted as a function of (i) activity for extensile and contractile suspensions, (ii) elasticity for extensile suspensions.}
\label{fig:ndefects}
\end{figure}

Having characterised the velocity field in terms of defects and jets we can motivate why the decay of the normalised velocity-velocity correlation function is independent of the activity. This follows because the defect density controls the spatial structure of the flow field, and hence the dependence of the velocity correlations on distance. Indeed, we find that the number of defects \cite{Huterer2005} in the simulation box depends only very weakly on the strength of the activity for fixed values of the elastic constant, as shown in Fig.~\ref{fig:ndefects}.

\begin{figure}
\includegraphics[trim = 40 100 50 25, clip, width=\linewidth]{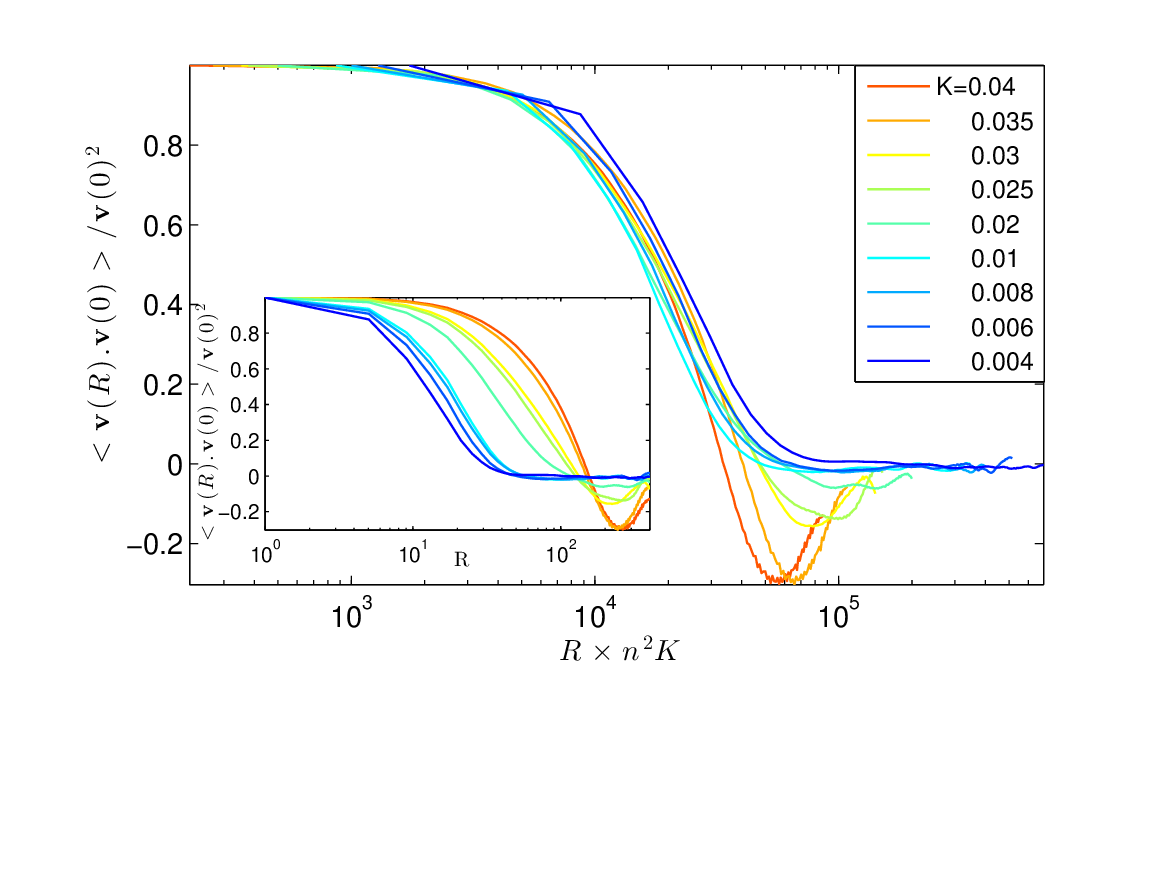}
\caption{Normalised velocity-velocity correlation functions as a function of scaled distance for different values of the elastic constant. Plots with unscaled distance are shown in the inset.} 
\label{fig:vcorrEl}
\end{figure}

To further investigate the relation between the defect density and the length scale of the flow field we also measured the number of point defects within the domain as a function of the elastic constant. Figure~\ref{fig:ndefects} shows that the number of defects increases sharply as the elastic constant is decreased. The normalised velocity-velocity correlation functions are plotted in Fig.~\ref{fig:vcorrEl}.  A collapse of the data is obtained when the distance is scaled by square of number of defects ($n$) times the elastic constant ($K$).

\begin{figure}
\subfigure[Extensile]{\includegraphics[trim = 60 50 30 20, clip, height=0.49\linewidth]{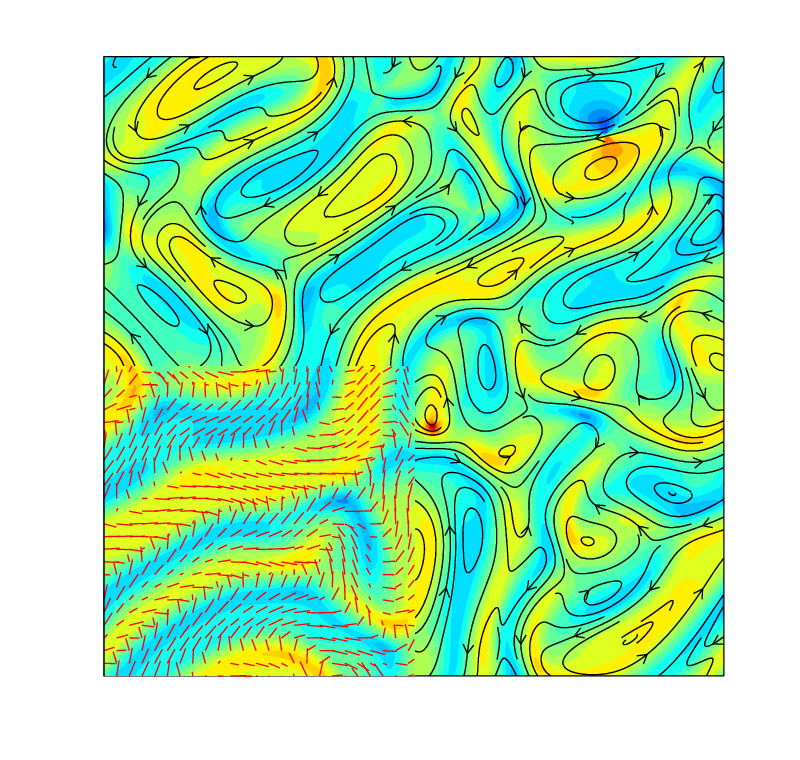}\label{fig:lowactext}}
\subfigure[Contractile]{\includegraphics[trim=60 50 30 20, clip, height=0.49\linewidth]{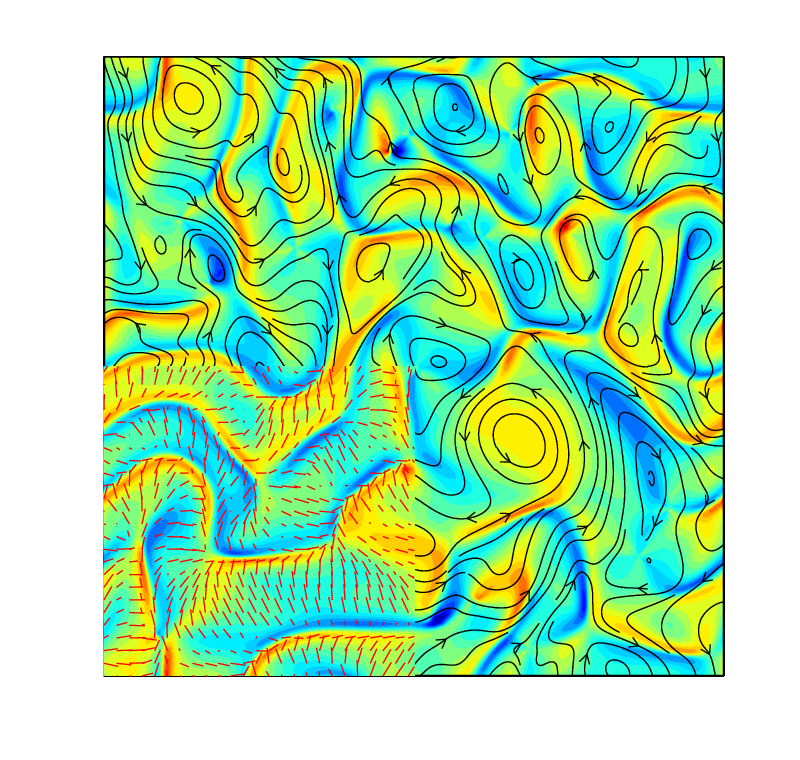}\label{fig:lowactcon}}
\caption{Patterns formed at low activity ($\zeta=0.001$) in (a) extensile and (b) contractile suspensions.  Colour shading indicates the vorticity field. The  director field is superimposed on one quarter of the plot, and streamlines are shown elsewhere. Lines of kinks are seen in the director field.}
\label{fig:lowact}
\end{figure}

Although the spatial scale of the flow is independent of activity, higher activity drives stronger jets \cite{Scott2009} and hence the RMS velocity of the flow field increases, approximately linearly with increasing activity. At very small activities  Fig.~\ref{fig:vcorrs} indicates that there is a regime where the flow velocity increases significantly more slowly. Similar behaviour is seen in the micotubule-kinesin experiments, although no comment  is made in the paper \cite{Dogic2012}. Moreover,  if concentration of swimmers may be considered as a measure of activity, a similar crossover was observed in particle simulations of extensile swimmers \cite{Shelly2012} where it was interpreted as a transition from a weakly correlated to strongly correlated dynamics of swimmers. In our simulations the low activity regime corresponds to the existence of lines of kinks in the director field, very similar to those observed in stable one-dimensional flows \cite{Davide2007}. These structures are illustrated in Fig.~\ref{fig:lowact} for an extensile and contractile suspension. The results suggest that such low values of the activity are sufficient to create, but not to further break up, the quasi-1D kinked director and flow configurations.

The results we observe are consistent with a simple scaling argument, based on the dynamics of the topological defects.
Balancing the rate of energy input via the active stress and the rate of viscous dissipation over a domain of size $\ell$ we find $v \sim \zeta \ell Q/ \mu$, which confirms our finding for the dependence of the velocity scale on the activity.

Moreover the rate of change of number of defects  may be written 
$\frac{dn}{dt} = \tilde{\alpha}-\tilde{\beta}n^2$
where  $\tilde{\alpha}$ is the rate of generation and $\tilde{\beta} n^2$ is the rate of annihilation of defects. Activity is responsible for the generation of defects, but at the expense of elastic free energy, so it is reasonable to assume $\tilde{\alpha} \sim \alpha \zeta / K$ where $\alpha$ is a proportionality constant. This scaling agrees with simulations tracking the rate of defect generation.
An expression for the rate of annihilation  may be constructed from kinetic theory. If $\sigma$ is the scattering cross section for defect annihilation, the rate of destruction is $\beta \sigma v n^2$ where $1/n \sigma$ is the mean free path and $\beta$ is a proportionality constant. Assuming that the defect velocity is of order the fluid velocity we can use the scaling argument given above to write $v \sim \zeta \ell Q/\mu$.

At steady state, the rate of creation and rate of destruction of a pair of defects are equal. Hence
$\alpha \frac{\zeta}{ K} = \beta \frac{ \sigma \zeta \ell Q n^2}{\mu}$
giving $\ell \sim 1/n^2K$. Therefore the relevant length scale characterising the velocity field is indeed independent of the activity. Moreover the dependence of $\ell$ on $n$ and $K$ gives the data collapse demonstrated in Fig. 4.

The dynamics of the director field and the velocity field are interconnected through a feedback loop. The hydrodynamic instabilities give rise to lines of distortions in the director field. At sufficiently high activities these are unstable to the formation of defect pairs which are driven apart by the active flow \cite{Giomiarxiv}. The
proliferation of defects, leads to a dynamic structure resembling the high temperature phase of a system with a defect-mediated phase transition, such as the Kosterlitz-Thouless transition\cite{Nelsonbook}.

A quantitative comparison of experiment and simulation needs us to choose a length scale $L$ and a time scale $T$. $L$ was measured from the decay of the $\langle vv \rangle$ correlations. $T$ was estimated by choosing the activity at which the transition in slope occurs in Fig.~\ref{fig:vcorrcom} as a reference value and, for this value, using passive scalar tracking to obtain the crossover time-scale from the ballistic to the diffusive regime. This gives $L=200 \mu m$ and $T=200s$ for the experiments (see Fig.~2 in \cite{Dogic2012}) and $100$ and $10^4$ lattice Boltzmann units for the simulations,  and hence characteristic velocities of $1 \mu m/s$ and $0.01$ LB units. These values are used to normalise the RMS velocities in Fig.~\ref{fig:vcorrcom} giving very similar numerical values. Thus, this study takes a step towards bridging the current gap between theory and experiments on active nematics. Both $1/2$ and $-1/2$ defects have been observed experimentally \cite{Dogic2012} and it would be interesting to relate their dynamics to the length scale set by the velocity correlations.  We also hope to motivate studies of active turbulence in other systems to look for similar properties of the flow field.

\begin{acknowledgments}
We are grateful to Z. Dogic for providing the experimental data in Fig.~1, and to M. Blow, J. Dunkel, M. Ravnik and a referee for helpful advice.  We acknowledge funding from the ERC Advanced Grant MiCE.
\end{acknowledgments}

\bibliography{refe.bib}

\end{document}